# CONSIDERATIONS ON THE PROCESS OF TARGET SELECTION FOR THE COMET INTERCEPTOR MISSION

C. Snodgrass[1*], E. Mazzotta Epifani[2], C. Tubiana[3], J. P. Sánchez[4], N. Biver[5], L. Inno[6,7], M. M. Knight[8], P. Lacerda[9,10], J. De Keyser[11,12], A. Donaldson[1], N. J. T. Edberg[13], M. Galand[14], A. Guilbert-Lepoutre[15], P. Henri[16,17], S. Kasahara[18], H. Kawakita[19], R. Kokotanekova[20,21], M. Kueppers[22], M. Micheli[23], M. Pajusalu[24], M. Rubin[25], N. Sakatani[26], K. Yoshioka[27], V. Della Corte[7], A. I. Eriksson[13], M. Fulle[28], C. Holt[29,30], L. Lara[31], A. Rotundi[32,6], E. Jehin[33]

[1]Institute for Astronomy, University of Edinburgh, Royal Observatory, Edinburgh EH9 3HJ, UK
[2]INAF - Osservatorio Astronomico di Roma, Via Frascati 33, 00040 Monte Porzio Catone (RM), Italy
[3]INAF - Istituto di Astrofisica e Planetologia Spaziali, Via del Fosso del Cavaliere 100, 00133 Roma, Italy
[4]Fédération ENAC ISAE-SUPAERO ONERA, Université de Toulouse, France
[5]LIRA, Observatoire de Paris, Université PSL, CNRS, Sorbonne Université, Université Paris Cité, CY Cergy Paris Université, 5 place Jules Janssen, 92190 Meudon, France
[6]Department of Science and Technology, Parthenope University of Naples, Centro Direzionale di Napoli, Naples, I-80143, Italy
[7]INAF, Osservatorio Astronomico di Capodimonte, Salita Moiariello, 16, Naples, I-80131, Italy.
[8]Physics Department, United States Naval Academy, 572C Holloway Rd, Annapolis, MD 21402
[9]Instituto de Astrofísica e Ciências do Espaço, Universidade de Coimbra, Portugal
[10]Instituto Pedro Nunes, Coimbra, Portugal
[11]Space Physics Division, Royal Belgian Institute for Space Aeronomy (BIRA-IASB), Ringlaan 3, B-1180, Brussels, Belgium
[12]Center for mathematical Plasma Astrophysics, Katholieke Universiteit Leuven, Celestijnenlaan 200B, B-3001, Heverlee, Belgium
[13]Swedish Institute of Space Physics, Uppsala, Sweden
[14]Department of Physics, Imperial College London, London SW7 2AZ, United Kingdom
[15]LGL TPE, UMR 5276 CNRS, Université Lyon 1, ENSL, 69622 Villeurbanne, France
[16]Laboratoire Lagrange, Observatoire Côte d'Azur, Université Côte d'Azur, CNRS, Nice, France
[17]LPC2E, CNRS, Université d'Orléans, CNES, Orléans, France
[18]The University of Tokyo, 7-3-1, Hongo, Bunkyo, Tokyo, Japan
[19]Koyama Space Science Institute, Kyoto Sangyo University, Motoyama, Kamigamo, Kita, Kyoto, Japan
[20]Institute of Astronomy and NAO, Bulgarian Academy of Sciences, 72 Tsarigradsko Chaussee Blvd., 1784 Sofia, Bulgaria
[21]International Space Science Institute, Hallerstrasse 6, 3012 Bern, Switzerland
[22]European Space Agency (ESA), European Space Astronomy Centre (ESAC), Camino Bajo del Castillo s/n, 28692 Villanueva de la Cañada, Madrid, Spain
[23]ESA NEO Coordination Centre, Planetary Defence Office, Largo Galileo Galilei, 1, 00044 Frascati (RM), Italy
[24]Tartu Observatory University of Tartu 61602 Tõravere Estonia
[25]Space Research & Planetary Sciences, Physics Insitute, University of Bern, Sidlerstrasse 5, CH-3012 Bern, Switzerland
[26]Institute of Space and Astronautical Science, Japan Aerospace Exploration Agency, 3-1-1 Yoshinodai, Chuo-ku Sagamihara, Kanagawa, 252-5210, Japan
[27]The University of Tokyo, 5-1-5, Kashiwano-ha, Kashiwa, Chiba, Japan
[28]INAF - Osservatorio Astronomico di Trieste, Via Tiepolo 11, 34143 Trieste Italy
[29]Las Cumbres Observatory, 6740 Cortona Drive Suite 102, Goleta, CA 93117, USA
[30]LSST-DA Catalyst Postdoctoral Fellow
[31]Instituto de Astrofísica de Andalucía (CSIC), c/Glorieta de la Astronomía 3, 18008 Granada, Spain
[32]UNESCO Chair "Environment, Resources and Sustainable Development," Department of Science and Technology, Parthenope University of Naples, Italy
[33]STAR Institute - University of Liège, Allée du 6 Août 19C, B-4000 Liège 1, Belgium

* e-mail: csn@roe.ac.uk

**Abstract**

Comet Interceptor is an ESA science mission with payload contributions from ESA Member States and with an international participation by JAXA. It is the first mission that is being designed, built, and potentially launched *before* its target is known. This approach will enable the spacecraft to perform the first mission to a Long Period Comet from the Oort Cloud, as these comets have fleeting visits to the inner Solar System lasting only months to years from first discovery, too short for the usual process of mission development to be followed. In this paper we describe a number of factors that need to be considered in selecting a target for the mission, including scientific, orbital, spacecraft and instrument constraints, and discussion of different prioritisation strategies. We find that, in the case where we have a choice of targets, our decisions will mostly be driven by orbital information, which we will have relatively early on, with information on the activity level of the comet an important but secondary consideration. As cometary activity levels are notoriously hard to predict based on early observations alone, this prioritisation / decision approach based more on orbits gives us confidence that a good comet that is compatible with the spacecraft constraints will be selectable with sufficient warning time to allow the mission to intercept it.

## 1. INTRODUCTION

Comets are often considered as the best source of information regarding how planets formed in the protoplanetary disc (Blum et al. 2017; Filacchione et al. 2024). However, comets are not all born equal: the established classification scheme (Levison 1996) of comets distinguishes between Ecliptic Comets (ECs), which have a quasi-correspondence with the group of Short Period Comets (SPCs), and Nearly Isotropic Comets (NICs), comprised of both the Halley-type comets and the Long Period Comets (LPCs). Among the latter, there are the subset of Dynamically New Comets (DNCs), typically with semimajor axis $a > 10^4$ au (also including comets on hyperbolic orbits), which are likely to be on their first passage into the inner Solar System. Since 2017, another class of cometary objects has been identified: the Interstellar Objects (ISOs), visitors coming from extrasolar planetary systems (Jewitt & Seligman 2023). 1I/`Oumuamua, 2I/Borisov, and 3I/ATLAS are the only members of this brand-new class identified to date. DNCs that are traveling towards the Sun for the first time are thought to be some of the most pristine remnants of planet formation that we can observe directly at present, having experienced no Solar heating since they were originally scattered into the Oort cloud as the planets formed. As such, they are a fundamental linkage between planetary systems, such as the Solar System, and our understanding of how protoplanetary discs form and evolve.

Only a handful of comets has been visited in-situ by space missions up to now (Snodgrass et al. 2024), and they are all SPCs (and all ECs apart from 1P/Halley, which still has a relatively short orbital period *P* of ~76 years): 26P/Grigg-Skjellerup (*P* ~5.3 years), 19P/Borrelly (*P* ~6.9 years), 9P/Tempel 1 (*P* ~5.6 years), 81P/Wild 2 (*P* ~6.4 years), 103P/Hartley 2 (P~6.5 years ), and 67P/Churyumov-Gerasimenko (*P* ~6.4 years), the target of the highly successful Rosetta mission. The repeated passages through the inner Solar System, at close heliocentric distances, and therefore the repeated exposures to the "ageing" action of the Sun, make all these comets *processed* targets to some extent, more than essentially *pristine* bodies. The next step in the field of cometary science will be the new European Space Agency (ESA) (in cooperation with the Japan Aerospace Exploration Agency (JAXA)) cometary mission Comet Interceptor (CI) (Jones et al. 2024), which aims to investigate in situ, for the first time, the nucleus and environment of a LPC, preferably a DNC, or even an ISO, thus a surely less evolved object than SPCs.

LPCs have historically been identified from a few years to a few months prior to their perihelion, too short a period to enable a mission to be planned and launched. Therefore, another novel element of the CI mission will be that it is being designed, realised and will potentially be launched *before* its target is discovered. CI could wait in a parking orbit around the Sun-Earth L2 point, where it can station-keep with very little fuel, until a newly-discovered reachable comet is selected. The spacecraft will depart L2 to encounter the comet at a distance from the Sun of ~ 1 au, with a cruise phase of up to 3 years. The intercept will involve a close-approach flyby scenario using three elements: a mother spacecraft, spacecraft A, and smaller probes named B1 and B2 that are carried as payloads until shortly before the flyby and delivered to different flyby trajectories. This will allow the gathering of remote and in situ multi-point observations of the comet and its coma.

Within this general mission scenario, the target comet must still be discovered inbound at a relatively large distance, to give sufficient time to characterise its orbit and activity levels, and for the spacecraft to reach the encounter position. This would have been impossible a decade ago, when LPCs were typically discovered only a few au from the Sun, less than a year from their perihelion passage. Now and in the near future, powerful sky surveys, such as the Legacy Survey of Space and Time (LSST) of the Vera C. Rubin Observatory, will be able to detect and identify comets at relatively large distances (beyond 10 au) from the Sun. It is therefore possible that multiple potential targets may be discovered. In that case, in order to maximize the success of the mission, the different targets must be evaluated considering the mission's scientific and technical requirements, derived both from spacecraft and instrument constraints. To prepare for this process, it is necessary to define criteria for assessing the relative merit of different candidates (if more than one is available).

In this paper, we present an example of the "decision process", performed in order to identify a prioritisation strategy aimed at selecting the best possible target for the CI mission. In section 2, we describe the general criteria for target selection, including the constraints given by the mission goals, and by measurement requirements of the probes and instruments onboard CI. In section 3, we present the database of historical comets we used to exercise the selection process. Section 4 describes two possible approaches to downsize the historical database and to identify "virtual targets" that could have been selected for a putative CI mission, if it had been already operating when these comets were accessible. The purpose of this paper is not to advocate for either of these prioritisation strategies, but to describe the approaches followed and the different results obtained by each strategy, as these illuminate the various trade-offs to be considered in selecting a real target. In section 5 we discuss the output of the exercise, highlighting the target parameters that were found to be the most important for decision making, and identifying the main scientific work that needs to be done in preparation for the actual target selection.

## 2. GENERAL CRITERIA FOR OPTIMAL SELECTION OF THE CI TARGET

CI will encounter a comet close to 1 au from the Sun, likely at (or close to) its perihelion. To enable spacecraft design, a reference scenario was adopted by ESA for the operational phase at ~1 au, defined by a target with a comet activity (dust and gas production) similar to 1P/Halley and nucleus diameter of 10 km, with the main probe passing at 1000 km distance from the nucleus and probes B1 and B2 targeted to 850 km and 400 km respectively (Kidger 2023, Jones et al. 2024). This should be technically feasible (e.g. in terms of the necessary dust shielding to protect the spacecraft) as ESA achieved a successful flyby of Halley with the Giotto mission at a heliocentric distance $r_h$ = 0.89 au in 1986 (e.g. Keller et al. 1987). As Halley is a relatively active comet (i.e. produces a lot of dust), and has a retrograde orbit that means a high encounter velocity $v$ ~70 km/s, this is seen as a 'worst case' from an engineering point of view, with many high-speed dust impacts on the spacecraft. The real target is unlikely to be so challenging, ensuring a technically feasible encounter if the spacecraft is built to withstand a Halley-like comet. The final close approach distances for the spacecraft and probes could be adapted depending on the activity level of the real comet (De Keyser et al. 2024a), and an 'Engineering Dust Coma Model' was constructed by the mission science team to inform scaling of flyby distances depending on comet activity levels and encounter velocity (Marschall et al., 2022). The spacecraft design therefore provides a limiting case for target selection – a close and high-speed flyby of a higher activity comet, e.g. comparable in activity level to Hale-Bopp, is ruled out. As such 'great comets' appear to be rare, this is unlikely to be constraining on target selection.

The following sections describe the constraints in addition to spacecraft survival that should be considered to make a decision between multiple possible comets, should a number of 'survivable' ones be discovered at the same time. The criteria we discuss could in principle be quantified and weighted according to a scheme along the lines suggested by Vigren et al (2023), who also considered the evolution of the selection criteria as the remaining mission time decreases. We here concentrate on the criteria themselves, leaving implementation of such or similar procedures for later consideration.

### 2.1 OPERATIONAL CONSTRAINTS FROM THE SPACECRAFT

CI is expected to be launched in 2029, by an Ariane 62, in shared launch configuration with M4 ARIEL as prime payload. It will be commissioned into an injection free transfer towards a large amplitude quasi-halo orbit near the Sun-Earth L2 point. Halo orbits are unstable orbit configurations in the vicinity of the Earth, where CI can remain as long as necessary (at the cost of small station-keeping manoeuvres), but also allow for rapid departures to intercept a comet, when the opportunity arises. Nevertheless, to limit operational costs, the mission lifetime is limited to 6 years (CI Project Team, 2020; Jones et al, 2024), including launch, commissioning and insertion into the Sun-Earth L2 orbit (2-3 months), as well as 6-months of post flyby spacecraft A operations for science data downlink.

| Parameter | Minimum | Maximum | Units |
|---|---|---|---|
| Heliocentric distance at encounter | 0.9 | 1.2 | au |
| Heliocentric distance during cruise | 0.85 | 1.2 | au |
| Geocentric distance at encounter | - | 2 | au |
| Sun-Earth-Comet angle at encounter | 5 - 10 | - | degrees |
| Solar aspect angle at encounter | 45 | 135 | degrees |
| Relative velocity at encounter | 10 | 70 | km/s |
| Total effective Δ$v$ (including L2 Halo orbit departure) | - | ~1.5 | km/s |
| Comet dust environment | - | Halley-like | - |

*Table 1: Spacecraft constraints on trajectories and comet activity level that limit target choice.*

CI will be equipped with a chemical propulsion system able to provide a propulsive Δ$v$ of at least 600 m/s for the heliocentric transfer phase (Jones et al, 2024). By timing the departure from the L2 Halo orbit correctly, this is expected to give a total *effective Δv* of around 1.5 km/s to intercept the comet (Sánchez et al, 2021). It follows from this Δ$v$ constraint, as well as given the costs associated with reaching out of the ecliptic plane, that the ascending or descending node of the incoming target are the most advantageous locations to intercept the comet (Perry, 2019). Moreover, during the cruise the spacecraft trajectory must remain within 0.85 < $r_h$ < 1.2 au and the interception point (the comet's ascending/descending node) must be within the range 0.9 to 1.2 au, due to both the Δ$v$ constraint (Sánchez et al, 2021; 2024) and thermal design limits (ESA CDF Team, 2019). Further to Δ*v* limits and thermal constraints, the intersection must avoid superior solar conjunction with less than 10 degrees of the Sun-Earth-target angle or 5 degrees for inferior solar conjunctions to prevent communication issues during the encounter (CI Project Team, 2020). CI will be capable of performing a comet flyby at distances from Earth of up to 2 au. Beyond constraints on the encounter location, there are also requirements on the relative characteristics of it, particularly velocity and direction. These include a relative velocity that must be between 10 to 70 km/s and a solar aspect angle (i.e., angle between the target to sun vector and the flyby relative velocity) between 45 to 135 degrees. These were selected to provide limits for spacecraft design while encompassing the majority of possible encounters with LPC-like orbits near 1 au from the Sun (Jones et al, 2024). These parameters are constraints primarily for engineering reasons (design of dust shields to cope with an expected range of impact energies, and constraints on solar panel orientation for power), but do also have some general science influence: as described in the next section, a slower encounter is generally preferable for science, as it increases the time for data collection within a given distance. There is also a preference for approach angles from the day side to maximise the nucleus illumination for the lower-risk pre-CA measurements. These spacecraft constraints on trajectories, and therefore target choice, are summarized in Table 1.

Finally, there is a preference for encounters that are as close to Earth as possible. This means a higher data downlink rate after the encounter, as well as (typically) a shorter cruise phase and/or lower Δ*v*, giving more margin. A comet closer to Earth will also normally be better placed for supporting ground-based observations, which can enhance the mission by providing contextual studies (e.g., Snodgrass et al. 2017). Observability during the encounter by Earth-based telescopes was considered during the prioritisation exercise but not used as a reason to select or de-select comets.

## 2.2 OPERATIONAL, SCIENTIFIC AND PERFORMANCE CONSTRAINTS FROM INSTRUMENTS ON BOARD

| Instrument | Type | Max. $v$ km s$^{-1}$ | Min. $Q_{gas}$ molec. s$^{-1}$ | Min. $Q_{H2O}$ molec. s$^{-1}$ | Dust/gas $Af\rho/Q_{H2O}$ | Other preferences |
|---|---|---|---|---|---|---|
| A CoCa | Vis. camera | low | - | - | low | Flyby distance needs to be compatible with phase angle coverage. |
| A MANiaC | Neutral mass spectrometer | 40 - 50 | $10^{28}$ - $10^{29}$ | $10^{28}$ | - | Flyby as close as possible. |
| A MIRMIS | IR Camera & spectrograph | low | 5 x $10^{28}$ | - | - | - |
| A/B2 DFP | Dust, fields & plasma package | 50 | $10^{28}$ - $10^{29}$ | - | low | Flyby on the sunward side. $r_h$ low. Min. $v$ and max. flyby distance depend on $Q_{gas}$. |
| B1 PS | Magnetometer & ion mass spec. | 50 | $10^{28}$ | - | < $10^{-25}$ | A larger nucleus is better. |
| B1 HI | UV camera | 50 | $10^{28}$ | $10^{28}$ | < $10^{-25}$ | A larger nucleus is better. |
| B1 NAC/WAC | Vis. cameras | 50 | $10^{28}$ | - | < $10^{-25}$ | A larger nucleus is better. |
| B2 OPIC | Vis. camera | - | high | - | - | Close flyby. A larger nucleus is better. |
| B2 EnVisS | Polarimeter | - | - | - | high | - |

*Table 2. Summary of preferences for target selection from all instruments (listed by spacecraft/probe). Numeric values given where possible, or 'high/low' when there is a preference but no limiting value provided.*

To fulfil the mission Science Objectives, a suite of 10 instruments have been selected to compose the CI payload: 4 onboard spacecraft A, 3 onboard probe B1, 3 onboard probe B2 (for details, see Jones et al. 2024). This section describes preferences and constraints from the instruments, from both a performance and an operability point of view, to assure the best science return from a hypothetical target candidate. The high-level input from each instrument team is summarized in Table 2. More detailed reasoning for these preferences is given below for some of the instruments and sensors:

The Mass Analyzer for Neutrals in a Coma (MANiaC) instrument on spacecraft A will measure the composition of the gas coma. For target selection it provides a constraint on gas production rate $Q$: the higher the $Q$, the better for the instrument performance, as the better signal-to-noise ratio it will achieve in the available measuring time. A gas production rate (irrespective of the gas) of $Q = 10^{26}$ molecules/s will be barely detectable even for the most dominant species, providing a lower limit for the preferred activity level of the target comet. To measure the D/H isotopic ratio, a water production rate $Q_{H2O}$ of at least a few times $10^{28}$ molecules/s will be required. This instrument will also get better results with a lower speed flyby, preferably below 50 km/s and ideally below 40 km/s.

The Cometary Plasma Light Instrument (COMPLIMENT) within the Dust, Field and Plasma (DFP) suite on spacecraft A measures the electric field, sub-micrometer cometary dust, and plasma environment around the comet and has preferences on (i) the outgassing rate, (ii) the flyby velocity, (iii) the cometary dust flux and (iv) heliocentric distance at the time of the encounter: (i) To improve signal-to-noise, the outgassing rate should be as large as possible (De Keyser et al. 2024b). (ii) To improve scientific return and reduce risks, the cometary flyby speed should be as low as possible. A lower speed increases the measurement spatial resolution for the given fixed time resolution of the instrument, decreases the risk that cometary dust impacts damage the COMPLIMENT sensors, and decreases the risk that secondary electron emission from cometary neutral impact on the spacecraft generates a spurious electron cloud around the spacecraft thus blinding COMPLIMENT and other plasma instruments (Bergman et al. 2023). (iii) To ensure integrity of the sensors, the cometary dust flux shall be small enough to avoid destructive impacts on the sensors during the flyby. (iv) All other things being equal, a further selection criterion would be the heliocentric distance of the encounter. For COMPLIMENT and other plasma instruments, it is important to intercept the comet as close to the Sun as possible (within the range accessible by CI) since (a) the gas production rate then tends to be higher, and (b) the ionization rate is larger, thus enlarging the comet magnetosphere and enhancing the science return.

The Comet Camera (CoCa – the high resolution imaging instrument on spacecraft A) has a preference on the flyby encounter velocity $v$: the slower the $v$, the better for the instrument performance, as this gives more time to collect images and will allow individual exposure times to be longer without smearing.

The B1 probe and its instrument suite (provided by JAXA) provided a merged list of preferences, covering spacecraft navigation and survival as well as instrument measurements, that could influence target selection strategy: the selected comet should have a total gas production rate $Q_{gas} > 10^{28}$ molecules/s, and preferably $Q_{H2O} > 10^{28}$ molecules/s for the Hydrogen Imager (HI) instrument; the flyby velocity should ideally be slower than $v = 50$ km/s; there should be no evidence that the comet is going to disrupt; to avoid an unexpectedly harsh dust environment there should be no evidence of frequent outbursts during the inbound orbit and the comet should not be too "dusty" – the latter is defined as a limiting dust-to-gas ratio of

log( $Af\rho$[cm]/$Q_{H2O}$ [molecules/s] ) < -25; for on-board navigation, the total apparent magnitude of the target as seen by the spacecraft after B1 separation should be brighter than 5th magnitude. This also translates into a preference for larger and/or more active comets.

The FluxGate Magnetometer on probe B2 (FGM-B2), part of the Dust, Field and Plasma (DFP) suite, has a preference on the encounter speed which depends on the gas production rate. One key science goal is to identify the location of the bow shock (formed for high enough outgassing (Koenders et al. 2013)). This requires that probe B2 is released before the inbound crossing of the bow shock. For an activity level similar to the Giotto flyby of 1P/Halley ($Q_{gas}$ = 7x10$^{29}$ molecules/s), the bow shock would be encountered at B2 deployment (currently expected to be 20h before Closest Approach (CA)) for $v$=16 km/s. FGM-B2 has 3h-calibration planned in the solar wind, hence upstream of the bow shock. Including this calibration period after B2 release would require $v$ > 20 km/s. The lower limit for $v$ depends on $Q_{gas}$, decreasing/increasing for lower/higher $Q_{gas}$ (De Keyser et al. 2024b). The operation requirement would relax if the period from B2 deployment to CA is increased.

Another FGM-B2 science goal is to detect the diamagnetic cavity. Its highly variable outer boundary was detected by Rosetta at cometocentric distances between 200 and 400 km at $Q_{gas}$ = (3-5)x10$^{28}$ molecules/s in the terminator plane (Goetz et al., 2016). Its detection by FGM-B2, onboard B2 whose CA is currently planned at 400 km, would require $Q_{gas}$ > 5x10$^{28}$ molecules/s. This value would decrease for reduced CA distances of B2. If this is not fulfilled, FGM-B2 may lose the detection of the diamagnetic cavity (assuming the outgassing is high enough to have it formed), but would still be able to provide key measurements for the other regions.

One of the important features of CI is its ability to perform multi-point measurements, to disentangle spatial and temporal variations. This is particularly valuable for the study of the interaction between the Sun, its solar wind, and the target comet. In addition to FGM-B2 there are magnetometers on spacecraft A (FGM-A) and on probe B1 (MAG). B1 will be released from spacecraft A first, followed by B2. B2 will make the closest approach to the nucleus, followed by B1, and spacecraft A will have the most distant flyby. A comet with a large outgassing rate would ensure crossing both the bow shock and the diamagnetic cavity with spacecraft A (Edberg et al 2024). In the case of a low flyby velocity, it is very unlikely that probe B1 is released downstream of the bow shock, even if probe B2 is (De Keyser et al. 2024b), so that at least two-point measurements are likely ensured. In the case of a target comet with a high outgassing rate, probabilities for probe B2 of crossing the diamagnetic cavity are substantial.

## 2.3 DYNAMICAL CONSTRAINTS

The mission goal is to encounter a LPC, preferably a DNC, or an ISO. The latter is expected to be unlikely, given the relative rarity of ISOs passing within a reachable range for the spacecraft and the likely short warning time, but even the remote possibility of encountering a body from another star system is scientifically exciting. The timescales whereby the objects of highest interest, ISOs and DNCs, can be identified as such are likely very different. For the first ISO discovered in our solar system (1I/'Oumuamua), observations on an orbital arc of about 2 weeks were sufficient to unambiguously verify a strongly hyperbolic trajectory and a high relative velocity with respect to the Solar System, giving solid evidence of its interstellar origin (Meech et al., 2017). The interstellar nature of the second and third ISOs was recognised even quicker than this (only a few nights of observation; Guzik et al., 2019; Seligman et al., 2025).

On the other side, once a new LPC is discovered, timescales to discriminate, based on astrometry, between being dynamically "new" (i.e., on its first passage inside our Solar System, directly coming from the Oort cloud) or "returning" are much longer. An accuracy of the order of 10$^{-5}$ is needed on its eccentricity for this. This typically translates into needing a ~1 year of arc to have an indication, and ~2-3 years of arc to be reasonably sure, consistent with the findings of Krolikowska & Dybczynski (2020). A comet's activity pattern could also indicate the likelihood of a comet being a DNC (Holt et al., 2024, Lacerda et al., 2025), but this would also take months to years of observations.

## 2.4 PREDICTABILITY OF COMETS

The main difficulty for an optimal target selection is connecting what is known about comets' properties when they are at $r_h \sim 10$ au (discovery distance) to what the comets properties will be at $r_h \sim 1$ au (CI encounter distance). In general, at large heliocentric distance the first parameter that gives some hints on a comet's properties is its appearance – as a bare nucleus, or with a compact but detectable coma, or with extended coma and tail(s). The detection of activity in comets at $r_h \sim 10$ au and beyond is currently quite rare (e.g., Mazzotta Epifani et al. 2010; Jewitt et al. 2019a, 2019b; Hui et al. 2019; Farnham 2021), and it is even rarer to have monitoring of a comet's activity, both for dust and gas, over large heliocentric distance ranges. The total brightness of distant cometary targets is measurable only with the largest telescopes, and recent works (e.g., Holt et al., 2024) confirm that real comets do not necessarily follow a simple brightening law with a single slope at all distances, and that one should be very careful in making a prediction based on, e.g, 6-months of data at large heliocentric distance. Just extrapolating the activity index (Whipple, 1992) from the measurements at large heliocentric distance to perihelion, might be misleading: e.g., the DNC C/2020 R7 (ATLAS) was monitored starting around $r_h \sim 7$ au pre-perihelion and first exhibited an activity increase while approaching the Sun, then plateaued at ~5 au (Lister et al., 2022). Therefore, it is quite difficult to make reliable predictions on how a putative target, ideally selected beyond 10 au, will behave at the time of CI encounter: the mechanisms driving stalling, disintegration (see e.g. Jewitt 2022), colour changes, and outbursts (and the possible differences among returning and dynamically new comets) are still poorly known. Holt & Snodgrass (2025) introduced an empirical description of the non-linear brightening of LPCs at large distances, which can give a better statistical prediction but is still limited when considering the behaviour of an individual comet. Full thermophysical models (e.g., Meech & Svoren 2004, Bufanda et al. 2023) can predict activity levels but require assumptions on many parameters that will not be known for a newly discovered comet. The "probabilistic tail model" of Fulle et al. (2022) that was applied to the interstellar comet 2I/Borisov (Cremonese et al. 2020) can, in principle, be used to investigate how comets evolve before the onset of water-driven activity, but also has its own built-in assumptions on how comets work. The difficulties in predicting cometary activity translate into a strong preference for a pre-perihelion encounter for CI. As well as lower uncertainty in the extrapolation of activity levels, this also minimises uncertainties in the spacecraft trajectory (and $\Delta v$ needed to make corrections on approach) due to unknown non-gravitational forces changing the comet's orbit, and the risk that the comet may disintegrate before CI arrives.

For the purposes of performing a prioritisation exercise, we collected a list of historic comets and all known information on all of them, but we note that most of the observed parameters (e.g. composition, activity levels) were measured as they passed through the inner Solar System, and would not have been known in advance. This was a deliberate choice for the exercise of prioritisation – to decide which comets we would have preferred with the benefit of hindsight, and then to think about how we could have made such a decision with the information that would have been available at the time the comets were discovered.

## 3. LIST OF HISTORICAL COMETS

As a starting point to learn how to draw an informed decision on the best target to select, we compiled a list of historical comets that would have passed within the space potentially accessible to CI, had it been operating at the time: if all these comets were to arrive at once in the 2030s, which would be the best to choose, given all the possible scientific, operational, and instrumental constraints above?

| Parameter | Minimum | Maximum | Units |
|---|---|---|---|
| Perihelion distance | - | 1.5 | au |
| Aphelion distance | 100 (or undefined) | - | au |
| Data-arc span | 30 | - | days |
| Total number of observations | 30 | - | - |
| Orbit fit quality code | - | 5 | - |
| Comet absolute magnitude (M1) | defined | - | mag. |
| Comet total magnitude slope (K1) | defined | - | - |
| Node crossing heliocentric distance | 0.7 | 1.5 | au |

*Table 3. Parameters used to select suitable comets from the JPL Horizons database.*

The list started with definition of the criteria listed in Table 3 to select known historical comets from the JPL Horizons database (https://ssd.jpl.nasa.gov/horizons/), where orbit fit quality codes range from 0 for most constrained orbits to 9 for the least constrained orbits, and the node crossing distance was calculated from the orbital elements available in the database. The limits on perihelion distance and node crossing were deliberately chosen to be a bit broader than the real mission limits of $0.9 < r_h < 1.2$ au at comet flyby, in order to increase the number of targets considered, and to avoid missing any interesting cases that could have been just outside of the nominal limits but perhaps possible following more detailed analysis. The result was a list of 132 comets, the earliest being comet C/1898 F1 (Perrine) and the most recent being comet C/2023 R2 (PANSTARRS). The majority in the list were discovered in the past couple of decades, in the era of large sky surveys (Figure 1). These comets make up the "historical database".

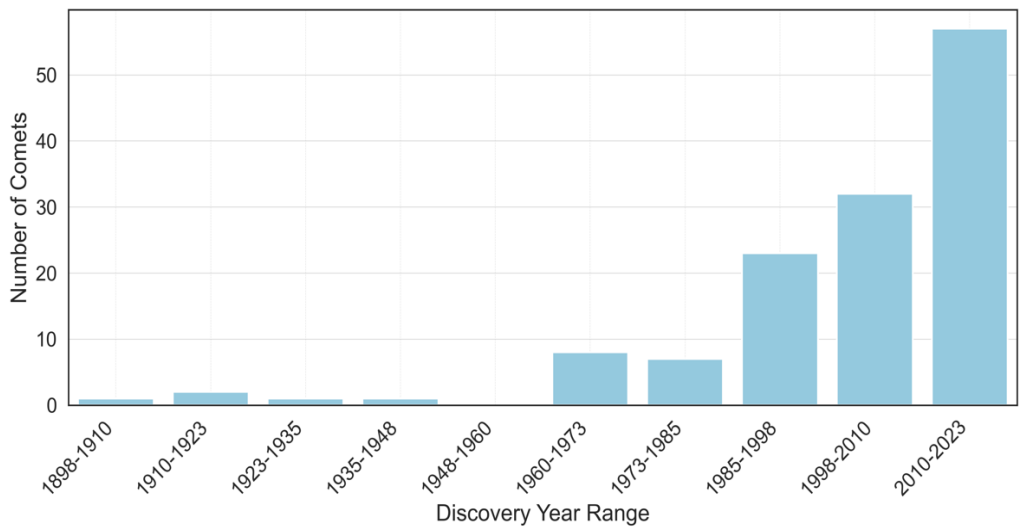

**Figure 1** – Year of discovery of the 132 comets of the "historical database".

For each comet in the historical database the following information was collected:
  A. (for all comets) orbit information from JPL Horizons; in particular, eccentricity *e*, perihelion distance *q*, inclination *i*, longitude of the ascending node, argument of perihelion, aphelion distance (where defined), time of perihelion, number of observations used. We also included the calculated heliocentric distance of the ascending and descending nodes in the database;

B. (for all comets) JPL fits to brightness evolution (absolute magnitude M1 and magnitude slope parameter K1); the use of the M1 magnitude as a proxy for the total brightness of the comet at 1 au is approximate, but generally OK as mostly these historic comets have been observed near 1 au. As stated above, extrapolation from early data for a comet discovered at a larger distance is riskier since real comets do not necessarily follow a simple brightening law with a single slope at all distances.

C. (for all comets) re-analysis of MPC data on magnitude and slope parameter and comparison with JPL Horizon data, separating inbound and outbound arcs and removing outliers. This more robust treatment confirms some trends and highlights limitations from the analysis from point B (Lacerda et al. 2025).

D. (for all comets) retrieval of the $1/a_0$ parameter (taken from the Nakano notes website: https://www.oaa.gr.jp/~oaacs/nk.htm[1]), where $a_0$ is the original semimajor axis prior to gravitational interactions in the inner solar system, to derive the level of "novelty" of a comet's passage in the Solar System. We used $1/a_0$ to estimate that the comets had orbits that were "hyperbolic/unbound" ($1/a_0 < 0$, 15 comets), dynamically new ($0 < 1/a_0 < 10^{-4}$ au$^{-1}$, 21 comets), "intermediate" ($10^{-4}$ au$^{-1} < 1/a_0 < 0.002$ au$^{-1}$, 21 comets), and returning ($1/a_0 > 0.002$ au$^{-1}$, 35 comets).

E. (for 24 comets) nucleus size, as derived by photometry, lightcurves, or estimates based on gas production rates (Groussin et al. 2010; Lis et al. 2019; Lamy et al. 2004; Bauer et al. 2017; Betzler et al. 2020, Jewitt 2022).

F. (for 31 comets) water production rate, in some cases measured both at perihelion and at ascending or descending node, from observations of H Lyman-α (Combi et al. 2019, 2021), OH in the UV (e.g., Schleicher et al. 2002a, 2002b; Opitom et al. 2015a, 2015b; Bair et al. 2018;) or radio (Crovisier et al. 2002), or $H_2O$ directly in the infra-red (Dello-Russo et al. 2016; Lippi et al. 2020, 2021, 2023) or sub-mm (Lecacheux et al. 2003, Biver et al. 2007, 2009, 2016, 2024).

G. (for 50 comets) information on coma abundances and composition, such as e.g. molecular ratios $HCN/H_2O$, $CO/H_2O$, $CH_3OH/H_2O$, $C_2/CN$, $C_3/CN$... (mostly taken from Robinson et al. 2024, collating literature values from Opitom et al. 2016, Bodewits et al. 2011, Dello Russo et al. 2016, Cochran et al. 2012, Fink 2009, Langland-Shula & Smith 2011, Lippi et al. 2021, Harrington-Pinto et al. 2022).

H. (for the 84 comets discovered from 2000 only, due to availability of suitably formatted orbit data) "feasibility study" for CI mission. A $\Delta v < 1.5$ km/s is considered feasible, based on the minimum fuel anticipated and the correct timing of departure from the parking orbit (Sánchez et al 2024). For feasible comets (35), the latest departure date and the relative flyby velocity has been computed, the latter being mostly dependent on orbital inclination: retrograde comets have faster encounter velocity. The heliocentric distance and the predicted JPL magnitude (the comets' predicted apparent visual total magnitude from simple extrapolation of the JPL M1/K1 fit to larger distances) at the latest departure date and 6 months before departure have also been considered for "feasible comets".

I. (for 11 comets) information on whether or not the comet survived perihelion passage (mostly from Sekanina, 2019, with a handful of other comets known to have disrupted noted). Comet disruption is generally unpredictable, but is observed to be more likely in comets with small perihelion distance and high M1, i.e., faint and, presumably, small ones (Bortle 1991). Jewitt (2022) also found that smaller comets were more likely to disrupt.

---

[1] $1/a_0$ values are available from three places: MPC, Nakano Notes, and the CODE Catalog (Krolikowska & Dybczynski 2020). The CODE Catalog has the most robust dataset, but is available for many fewer objects since it requires long orbital arcs (typically at least 2 years). In order to get $1/a_0$ values for as many of our objects as possible under the same system, we elected to use Nakano Notes, which Holt et al. 2025 found to better replicate the CODE Catalog than the MPC.

Figures 2 and 3 illustrate some parameters for the historical comets described above, derived from the JPL Horizon database. The absolute magnitude distributions in Figure 2 show a skew towards brighter comets, which is likely due to discovery biases, but it is notable that this persists when only considering comets found in the modern (post-2000) survey era. This implies that when restricting our analysis to only those comets with feasibility analysis (i.e., the post-2000 ones) we do not sample a very different population. With a couple of notable exceptions (C/1995 O1 (Hale-Bopp) and C/2007 E2 (Lovejoy)), the distribution of measurements of $Af\rho/Q_{H2O}$, i.e. the measured dust-to-gas ratio in the coma, is tightly clustered around the median $1.37 \times 10^{-26}$. This means that the vast majority of these comets are compatible with the B1 probe's preference to avoid a very dust-rich comet (see section 2.2), although it is worth noting that some of the brightest comets observed are outliers.

## 4. CRITERIA FOR TARGET SELECTION

The primary goal of CI is to study a new(er) comet, i.e., a body which is (more) pristine compared to an object that passed through the inner solar system several times. Reaching this goal (based essentially on an optimal target selection) will enormously increase the scientific output of the mission and will give breakthrough results in comet science. Following this primary goal, on the basis of the "historical database" described above, we defined two alternative approaches on the selection criteria for the purposes of our exercise:
- "science first" approach: following this idea, the historical database was screened on the basis of the known properties of each comet, and on what could have been extrapolated at the time of encounter, to leave a shortlist of preferred comets that could then be investigated as feasible/not feasible targets for the mission.
- "feasibility first" approach: following this idea, the historical database was screened first on the basis of "feasibility" to consider only comets that could be reached by CI, based on a simplified mission analysis (see point H in the description of the database above), and only after that on "scientific" priorities. It should be noted that this simple analysis of feasibility does not guarantee (or rule out) the possibility that the real mission could have reached these comets, but is a useful approximation. For real targets being considered, ESA will perform a detailed mission analysis as a prerequisite to the final target selection.

Both strategies allowed us to narrow down the historical database by removing targets that were considered less favourable, giving a shortlist to be considered in more detail. It should be noted that both approaches actually mix both scientific and operational constraints to make the selection, but the broad descriptions of "science first" vs "feasibility first" are used as helpful shorthand to label them. "Scientific" constraints include information on the orbital parameters of the comet, which will be known at least approximately at discovery, and on the physical parameters, which will very likely require further characterisation observations after the discovery. The real mission target selection will be done by considering any potential target individually, rather than following either of these strategies: the point of this exercise was to identify the most critical parameters to consider and the effects of making different cuts to our initial database.

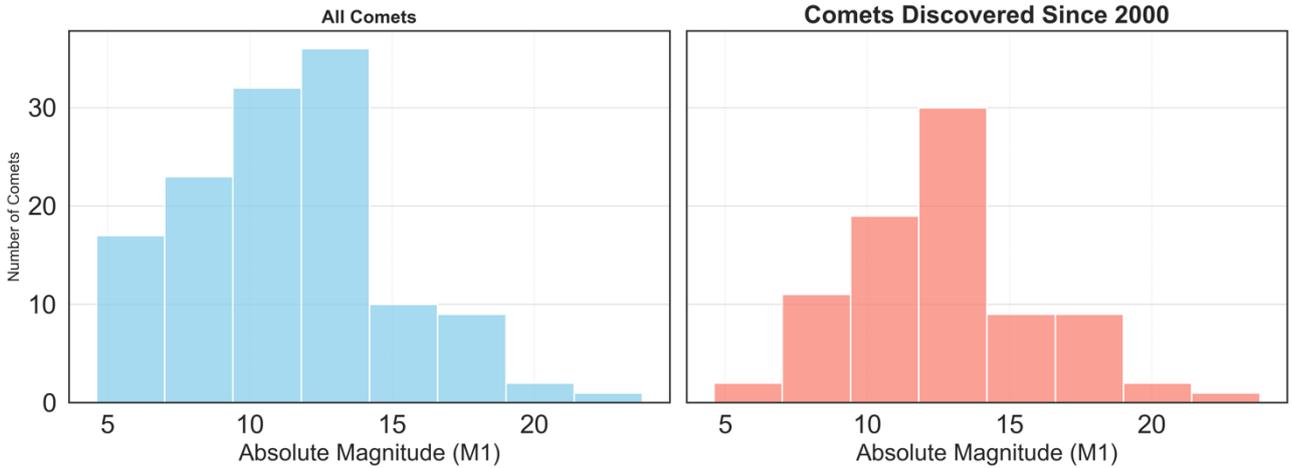

**Figure 2** - Distribution of absolute magnitudes M1 (as derived from JPL Horizon database) for all comets in the historical database *(left)* and for the subset of recent discoveries, i.e. since the year 2000 *(right)*.

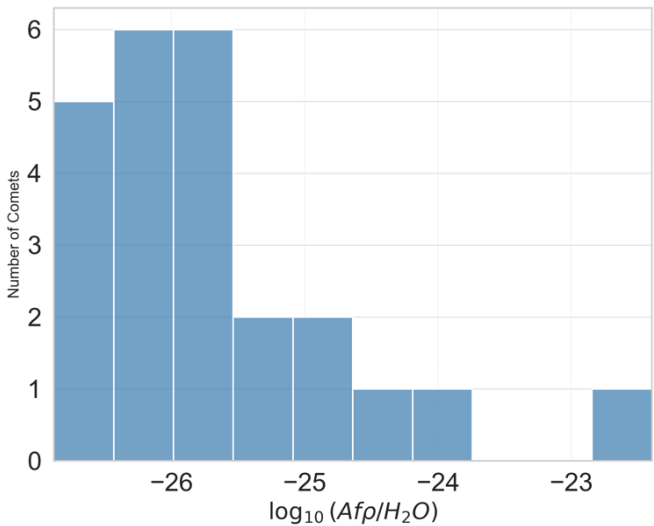

**Figure 3** – Information on the observed dust-to-gas ratio available for comets in the historical database.

### 4.1 APPROACH 1 – SCIENCE FIRST

Starting from the list of 132 historical comets, the following subsequent steps have been undertaken:
1) Cut everything that is not dynamically new / unbound or is unknown. **47 comets left**.
2) Filter away everything for which the encounter is outbound (only comets with encounter before perihelion are kept). **37 comets left**.
3) Filter by inclination of the orbit, to remove everything that is retrograde ($i > 90$ deg) and therefore has a higher encounter velocity. **23 comets left**. As the inclination distribution of new comets is not more isotropic than that of intermediate or old comets (Lacerda et al., 2025), this criterion does not impact the odds of picking a DNC.
4) Filter on brightness, as a proxy for total activity level (assuming similar dust/gas ratios, and that brighter comets have more of both), with brighter (more active) comets preferred for in situ measurements. The important question in this step is which is the best parameter for brightness, i.e. if the M1 parameter from JPL could be considered a reliable proxy of intrinsic brightness. Lacerda et al. (2025)'s recalculations based on MPC data were used to provide different levels of "goodness" based on the quality of data used to retrieve the value. This step demonstrates a crucial, still open question on how accurately we can predict a comet's brightness, and consequently how reliably we can set a brightness-based selection criterion. The adopted criterion was to remove comets which are fainter than JPL M1 = 10 mag (the exact value for the cut could be refined in future studies). **9 comets left**. It is worth mentioning that with this criterion C/2004 Q2 (Machholz) and C/2013 R1 (Lovejoy) were removed from the list despite being quite active with known (high) gas production rate, but faint M1, demonstrating a weakness in the assumption of equal dust/gas ratios.

The following criteria for further reduction were considered but not applied:

5) Gas production rate (or better, the availability of gas production rate information). However, this information will not be available for many (if any) comets at the distance/time of decision. In particular, no water production rates will be known at $r_h \sim$ 10 au when the decision could be expected to be taken, assuming a choice between comets with many years warning time pre-encounter.

6) Composition, where there is no clear preference for one type or another, so this isn't a very helpful criterion. It is also important to remember here that very likely no comparable composition information will be available when the actual selection decision shall be done. The expectation is that there will be a minimum of 6 months between discovery and departure. This should give some time to make observations at large telescopes to acquire information about composition and outbursts at large heliocentric distance (which is more a safety than a science requirement).

7) Feasibility, meaning that this comet could be reached by CI according to a simplified mission analysis, based on the minimum available $\Delta v$, calculated using the approach described by Sánchez et al 2021. Unfortunately this analysis was not available for all the remaining comets. Incidentally, for 5 out of the 9 remaining comets that have been investigated in this sense, only one is feasible.

It is worthwhile to underline that the first cuts were done on orbit parameters, and then cuts had to be done on observables (e.g. M1 magnitude), with the problem that it is still not clearly established which of those measured parameters are reliable and which are not. The first three cuts, on orbit parameters alone, are based on information that will be likely known soon after discovery (with the caveat that the question of new/returning may take longer to assess, as discussed in Section 2.3)

The remaining 9 comets flagged after this approach were sorted on brightness. Table 4 summarises their main orbital parameters and, when available, some information on physical properties and feasibility. The brightest comet in the sample is C/2011 Q2 (McNaught), a comet flagged as dynamically "unbound", which on the other hand has no information available on physical properties (nucleus size and composition) and gas production rate, and moreover is flagged as "unfeasible" from the mission point of view, due to its node crossing being almost 1.5 au, well outside the nominal 1.2 au limit reachable with conservative estimates about the available fuel. By chance, most of the 9 comets have relatively large node crossing distances (Table 4) and are not feasible for this reason. For less than half of the comets in the sample there is some physical information available, and the only "feasible" comet – **C/2003 T4 (LINEAR)** – is not among them. This latter comet is an "intermediate" target ($1/a_0 = 3.4 \cdot 10^{-4}$ au), which was discovered on 15-Oct-2003 when it was at $r_h$ = 6.5 au on its inbound orbital branch (McGaha et al. 2003).

| Comet | e | q [au] | Oort group | $i$ [°] | Node $r_h$ [au] | M1 | Info on composition | Feasibility |
|---|---|---|---|---|---|---|---|---|
| C/2011 Q2 (McNaught) | 1.0001 | 1.35 | UNBOUND | 36.87 | 1.48 | 4.7 | | Not feasible, node out of range |
| C/1978 H1 (Meier) | 1.0008 | 1.137 | New | 43.76 | 1.40 | 5.1 | Yes | Not tested, but likely not feasible, node out of range |
| C/1915 C1 (Mellish) | 1.0003 | 1.005 | Intermediate | 54.79 | 1.46 | 7.6 | | Not tested, but likely not feasible, node out of range |
| C/2021 S3 (PANSTARRS) | 0.9999 | 1.32 | New | 58.53 | 1.32 | 8.1 | | Not feasible, node out of range |
| C/1985 R1 (Hartley-Good) | 0.9999 | 0.695 | Intermediate | 79.93 | 1.32 | 8.5 | Yes | Not tested, but likely not feasible, node out of range |
| C/2014 Q2 (Lovejoy) | 0.9978 | 1.29 | Intermediate | 80.3 | 1.31 | 9.1 | Yes | Not feasible, node out of range |
| C/2003 T4 (LINEAR) | 1.0005 | 0.85 | Intermediate | 86.76 | 0.85 | 9.2 | | Feasible, $v$ = 55 km/s (but $r_h$ too low) |
| C/2014 Q1 (PANSTARRS) | 0.9997 | 0.315 | Intermediate | 43.11 | 1.26 | 9.7 | | Not feasible, node out of range |
| C/1995 Y1 (Hyakutake) | 1.0003 | 1.055 | UNBOUND | 54.47 | 1.25 | 9.9 | Yes | Not tested, but likely not feasible, node out of range |

**Table 4** - Summary of the main orbital parameters and, when available, information on physical properties and "feasibility" (based on a simplified analysis) of the 9 comets that were identified using approach 1 for the target selection criterion, sorted by JPL M1 absolute magnitude.

| Comet | e | q [au] | Oort group | $i$ [°] | Node $r_h$ [au] | M1 | Info on composition | Feasibility |
|---|---|---|---|---|---|---|---|---|
| C/2001 Q4 (NEAT) | 1.0007 | 0.962 | New | 99.64 | 0.96 | 8 | Yes | Feasible, $v$ = 57 km/s |
| C/2008 A1 (McNaught) | 1.0002 | 1.073 | New | 82.55 | 1.08 | 8.1 | | Feasible, $v$ = 46 km/s |
| C/2013 US10 (Catalina) | 1.0003 | 0.823 | New | 148.88 | 0.85 | 8.4 | Yes | Feasible, $v$ = 76 km/s (but too fast, and $r_h$ too low) |

**Table 5**- Summary of the main orbital parameters and, when available, information on physical properties and "feasibility" (based on a simplified analysis) of the 3 comets that were identified using approach 2 for the target selection criterion, sorted by JPL M1 absolute magnitude.

### 4.2 APPROACH 2 – FEASIBILITY FIRST

Starting from the list of 132 historical comets, the following subsequent steps have been undertaken:
1) cut on feasibility: remove all comets that are either not feasible or those whose feasibility had not been tested (as they were discovered before the year 2000). **28 comets left**.
2) cut on class: remove all comets that are not dynamically new (or for which this parameter is unknown). **17 comets left**.
3) Cut on orbit: among the remaining comets, the preference is for "new + unbound" rather than "intermediate". **6 comets left**.
4) Cut on brightness. The same criterion of strategy 1 has been adopted to remove comets which are fainter than 10 mag based on M1 from JPL (the exact value for the cut shall be refined). **3 comets left**.

The remaining 3 comets flagged after this approach are therefore sorted on brightness. Table 5 summarises their main orbital parameters and, when available, some information on physical properties and feasibility. The M1 parameter is rather faint (~8) for all the 3 comets, quite comparable to the M1 value for the only feasible comet obtained as output from approach 1;

the brightest target of the sample – **C/2001 Q4 (NEAT)** – is a new comet ($1/a_0$ = 6.0 · 10$^{-5}$ au), which was discovered on 24-Aug-2001 when it was at $r_h$ = 10.1 au on its inbound orbital branch (Pravdo et al. 2001). The other two comets – **C/2008 A1 (McNaught)** and **C/2013 US10 (Catalina)** – are also new comets ($1/a_0$ = 1.0 · 10$^{-4}$ au and $1/a_0$ = 5.23 · 10$^{-5}$ au, respectively). They were discovered on 2008-Jan-01 at $r_h$ = 3.74 au on its inbound orbital branch (McNaught et al. 2008) and on 2013-Oct-31 at $r_h$ = 8.31 au on its inbound orbital branch (Honkova et al. 2013), respectively.

The two approaches do not come to the same conclusions on the best comets as the 'best' ones from the 'science first' approach were mostly not feasible, so were rejected at step 1 of the 'feasibility first' approach. This was a result of our deliberate choice to consider a wider range of node distances (and therefore more real comets) for this exercise than are feasible for the nominal fuel load of the real mission. The three best ones from the second approach were rejected at steps 2 and 3 of the 'science first' approach, because they were either post-perihelion encounters (C/2008 A1) or retrograde orbits (C/2001 Q4 and C/2013 US10). It is worth noting that C/2013 US10, while 'feasible' in terms of being reachable within the *Δv* and time-of-flight limits of the simplified analysis, would have an encounter velocity (76 km/s) outside the spacecraft limits (70 km/s). Both C/2013 US10 and C/2003 T4 (the 'feasible' comet from the 'science first' approach) have node crossings (and therefore encounters) at $r_h$ = 0.85 au, closer than the expected 0.9 au limit. Whether or not these comets could actually be reachable by CI would need to be studied carefully by ESA. This illustrates the importance of considering each of the parameters we used in the two approaches individually for any real potential target; few comets are likely to have ideal values in all categories and trade-offs will have to be made. The point of this exercise was to identify what these key parameters and trade-offs are.

## 5. DISCUSSION

The two different approaches described in the previous section gave as output four comets (one from approach 1 and three from approach 2), which we defined as "preferred historical targets". This doesn't mean that only 4 comets in 40 years would have been available for a mission like CI, but that these are the best ones for the mission that resulted from a prioritisation exercise based on our chosen criteria, from the subset of those which have enough historical observations to apply our criteria to filter. The goal of the exercise was not to get the statistics of how many potential targets have been observed but rather to study the criteria needed to make the CI target decision.

For each of the four comets, we can compare their measured properties with the mission and instrument requirements and preferences listed in Tables 1 and 2. The compatibility with the mission $r_h$ and *v* requirements was discussed above. Only C/2008 A1 satisfies the desired *v* < 50 km/s limit from the instrument teams, although C/2003 T4 and C/2001 Q4 were not far beyond it (55 and 57 km/s, respectively). Water production rates were measured near $r_h$ = 1 au for all four and range from $10^{28}$ to 2.4 x $10^{29}$ molecules s$^{-1}$, meeting the requirements for in situ observations, with C/2001 Q4 having the highest rate. Only C/2001 Q4 has a measured dust-to-gas ratio, with *Afρ/$Q_{H2O}$* = 1.3 x 10$^{-26}$ (Fink 2009), compatible with the limits on this parameter. This ratio and gas production imply an *Afρ* ~ 3000 cm, or an order of magnitude lower dust production than Halley, suggesting that a closer than nominal flyby would have been possible for this comet; this would likely be limited by the maximum angular rate of the tracking mirrors rather than concerns about dust impacts. Nucleus radii were estimated from observations for C/2008 A1 (~4 km; Bauer et al. 2017) or from models for C/2001 Q4 and C/2013 US10 (2.7 and 2 km, respectively; Jewitt 2022). For the nominal flyby distances of the three spacecraft and probes, these would give images at closest approach with at least 500, 250, 64, 22 and 15 pixels across the nucleus for CoCA, NAC, OPIC and the short and long wavelength channels of MIRMIS, respectively.

Regarding feasibility of a mission to these targets, Figures from 4 to 7 summarise data on each comet's position on its orbital arc at discovery date, latest departure date (Ldd) and some critical dates before the latter: 6 months, 1, 2, and 3 years before Ldd. This is needed to derive where (and with which brightness) the comets should have been discovered and from when they should have been characterised, in anticipation of an encounter, to have enough time to discriminate its nature (interstellar, new, or returning), derive its main characteristics, and make reliable predictions on their evolution until the encounter.

It is notable that given the typical cruise durations of 1-3 years, the comets are often still at large distance from the Sun at the Ldd. Three of the four preferred historical targets would require a departure when the comet was still well beyond the water ice sublimation region, and the fourth was still at 4 au so barely in this region, so it would be impossible to measure $Q_{H2O}$ directly before a decision would have had to be made. While high gas production rate is a clear preference for most science measurements by the mission, we will only (possibly) have detections of more volatile species like CO and $CO_2$, and will need to use models to extrapolate to total (water dominated) gas production rate at the location of the encounter, near $r_h$ = 1 au. Direct observation of primary volatiles in comets at large distance is now possible with JWST, but the example of C/2024 E1 (Wierchzos), discovered at 8 au after our exercise was conducted, and observed by JWST as a "CI-like" comet that will reach a perihelion inside 1 au, shows that water production was not detected at $r_h$ = 7 au (Snodgrass et al. 2025). Typically, a decision on a target would need to be made with the comet further away than that. Our historical cases show that the brightness evolution (characterised by predicted M1 magnitude) does not always predict the water production rate well, although on average these do correlate (Jorda et al. 2008).

We also note that the use of the JPL M1 magnitude to represent total magnitude at 1 au is a simplification. Lacerda et al. (2025) showed that, for well observed comets in the modern survey era, reliable predictions can be found from the diverse photometry submitted to the Minor Planets Center, but there are significant outliers, and the automatic JPL fit to this photometry does not perform the careful selection and weighting that is required. The value of the M1 parameter listed in Tables 4 and 5, taken from the JPL database, does not always accurately reflect the total brightness of the comet. It should be noted that this discrepancy does not affect the discussion on the criteria to be used for selecting the target. A different parameter to describe the expected brightness of the comet (e.g. the heliocentric average total magnitude measured at 1 au) might be more suitable and more reliable but was not available for all comets in our historic database. The predicted / extrapolated JPL magnitudes at critical dates defined above are summarised in Table 6 but are subject to the same caveats. Using the approach of Inno et al. (2025), which takes these extrapolations at face value, all of these comets would have been detected by LSST (which has a limiting magnitude of 24.5) at least a year before Ldd and would therefore be feasible from the point of view of being discovered in time to consider them.

|  | **C/2003 T4** | **C/2001 Q4** | **C/2008 A1** | **C/2013 US10** |
|---|---|---|---|---|
| Discovery | 6.5 au, 18.7 | 10.1 au, 19.7 | 3.7 au, 16.3 | 8.3 au, 18.8 |
| Latest departure date (Ldd) | 6.3 au, 18.4 | 4.2 au, 15.4 | 7.9 au, 21.6 | 8.7 au, 18.9 |
| Ldd – 6 months | 7.8 au, 19.9 | 6.0 au, 17.2 | 9.3 au, 22.7 | 10.1 au, 20.1 |
| Ldd – 1 year | 9.3 au, 20.4 | 7.6 au, 18.4 | 10.7 au, 23.5 | 11.4 au, 20.3 |
| Ldd – 2 years | 11.9 au, 21.7 | 10.4 au, 20.0 | 13.1 au, 24.9 | 13.8 au, 21.3 |
| Ldd – 3 years | 14.3 au, 22.7 | 12.9 au, 21.1 | 15.4 au, 25.9 | 16.0 au, 22.1 |

**Table 6** – Heliocentric distances and JPL visual magnitude of the "preferred historical targets" at critical dates: at discovery, at the latest departure date given by feasibility analysis (Ldd), at Ldd minus 6 months, minus 1, 2, and 3 years before Ldd.

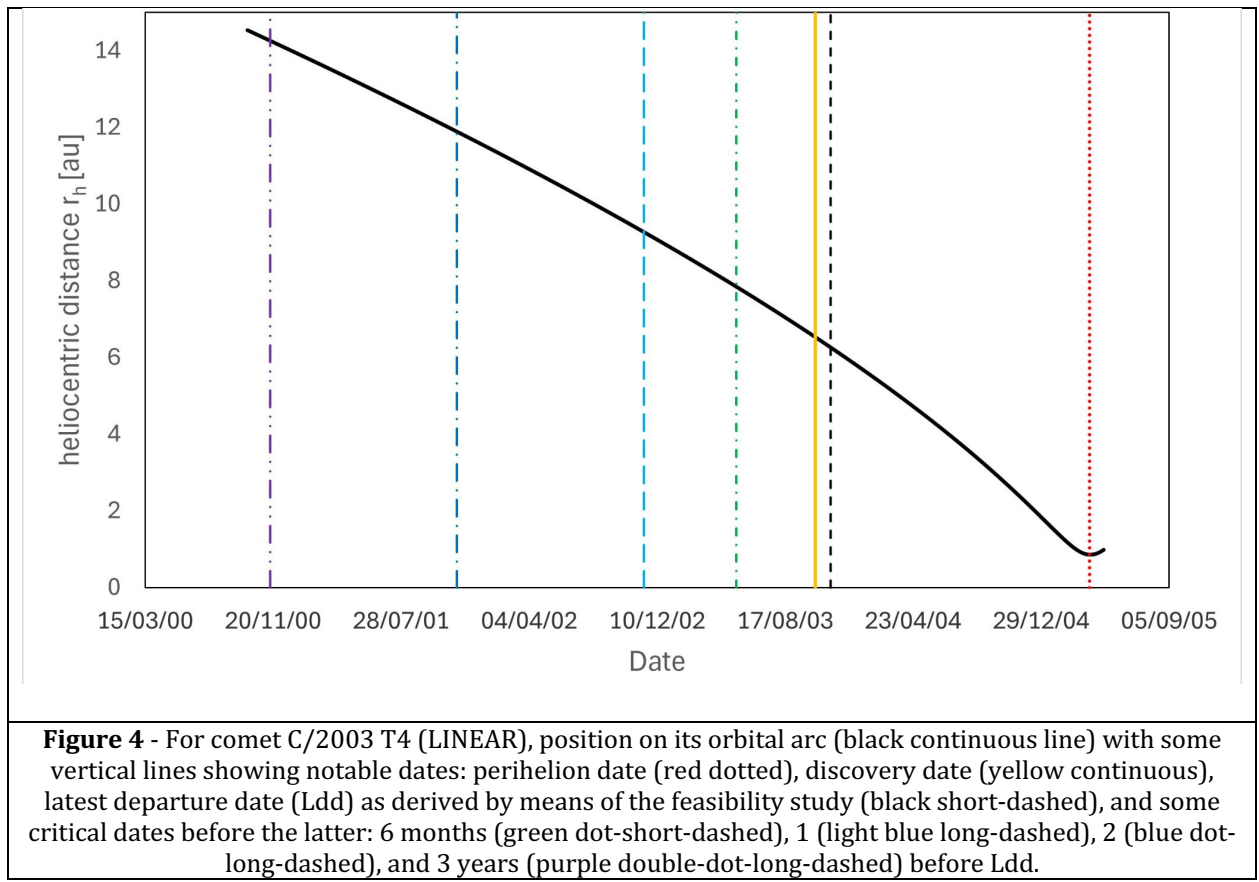

**Figure 4** - For comet C/2003 T4 (LINEAR), position on its orbital arc (black continuous line) with some vertical lines showing notable dates: perihelion date (red dotted), discovery date (yellow continuous), latest departure date (Ldd) as derived by means of the feasibility study (black short-dashed), and some critical dates before the latter: 6 months (green dot-short-dashed), 1 (light blue long-dashed), 2 (blue dot-long-dashed), and 3 years (purple double-dot-long-dashed) before Ldd.

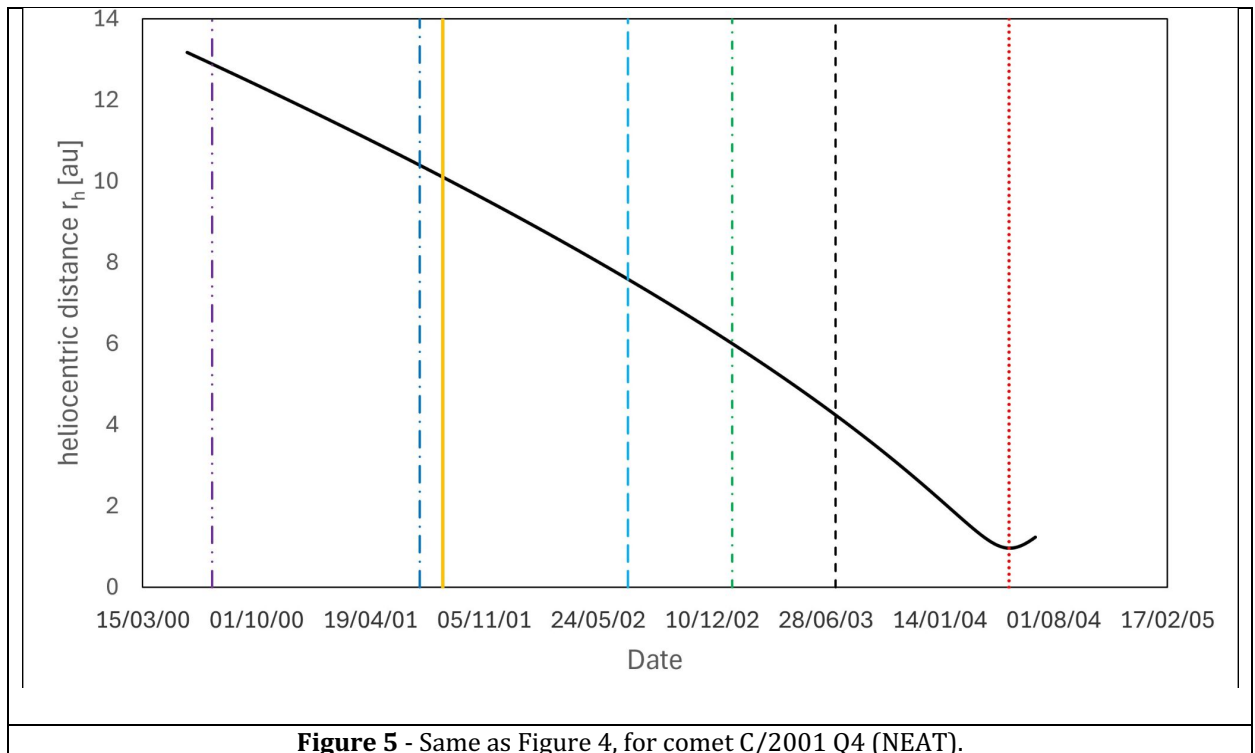

**Figure 5** - Same as Figure 4, for comet C/2001 Q4 (NEAT).

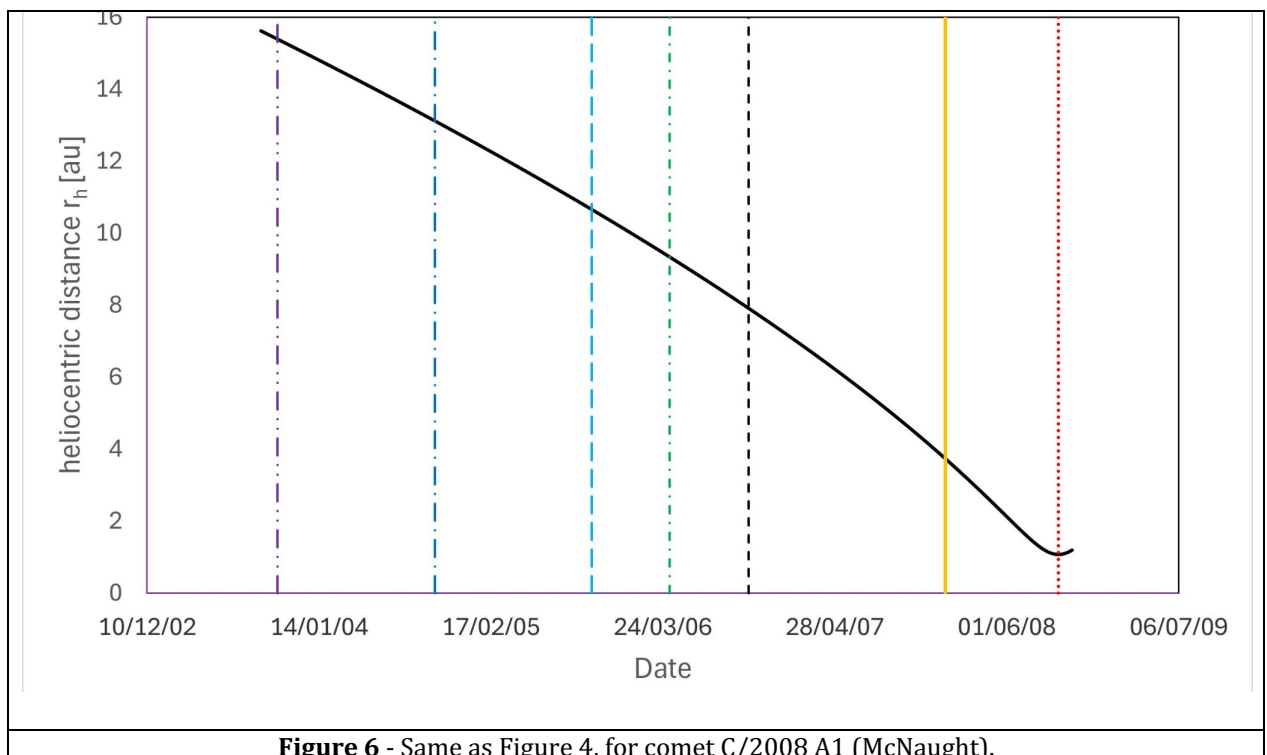

**Figure 6** - Same as Figure 4, for comet C/2008 A1 (McNaught).

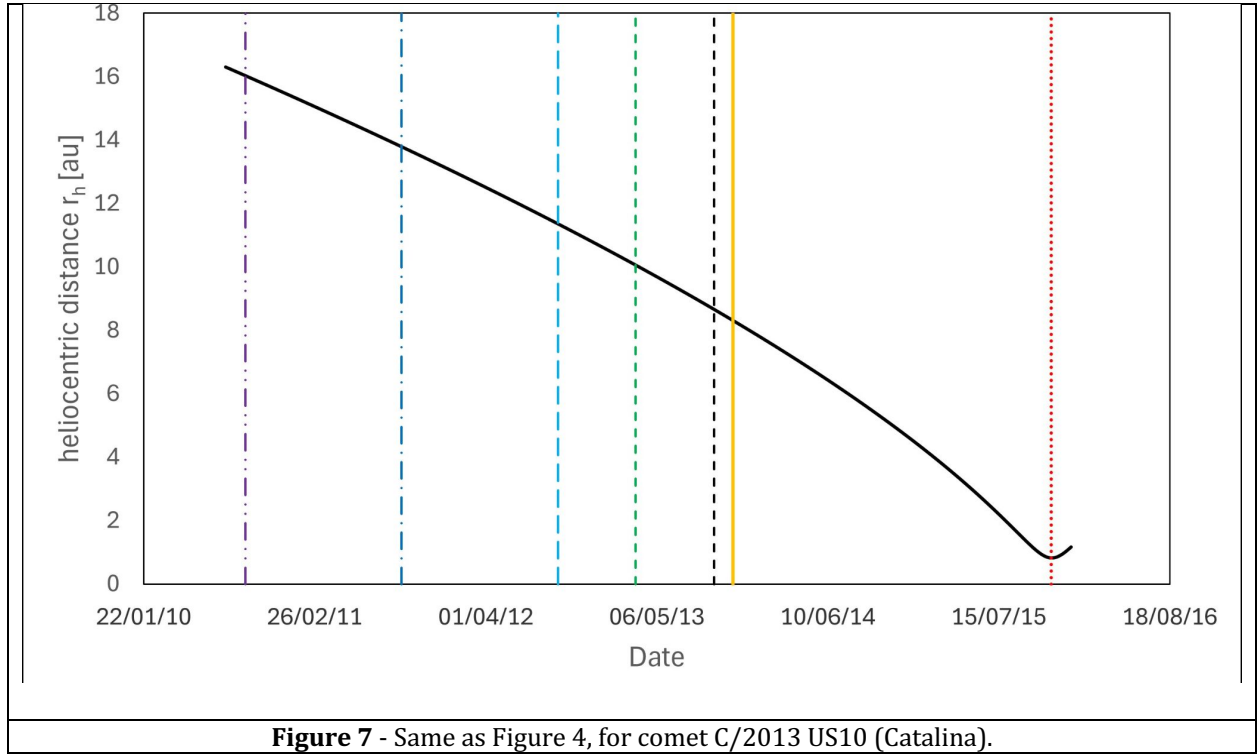
**Figure 7** - Same as Figure 4, for comet C/2013 US10 (Catalina).

Comet C/2001 Q4 (NEAT) is the only preferred historical target for which a CI-like mission could have been possible without benefiting from today's observing facilities: it was discovered when it was beyond 10 au, on its inbound orbital branch towards its perihelion $q$ that would have occurred 2.75 years after, on 2004-May-15, at 0.96 au. With a latest departure date on 2003-Jul-15, ~2 years would have been available to make an informed decision on a "go/no go" option.

The dynamical nature of the comet, as a "new target" coming from the Oort cloud and probably on its first passage within the inner Solar System, was secured a few months after discovery (Marsden 2002) and confirmed following longer-term monitoring of its orbit and calculation of the original $1/a_0$ (Marsden 2004). Physical characterisation (imaging and spectroscopy) started in mid-March 2002, when the comet was at 8.6 au (7 months after discovery, 1.3 years before Ldd) (Tozzi et al. 2003). Its close perihelion passage with good observing conditions from Earth (geocentric distance $\Delta$ ~0.4 au and solar elongation $\varepsilon$ ~70°) allowed to study in detail its coma composition (Wooden et al. 2004; Kawakita et al. 2005, 2006; Friedel et al. 2005; Remijan et al. 2006; Milam et al. 2006; Biver et al. 2009; Shinnaka et al. 2012; Rousselot et al. 2012; Ivanova et al. 2013; de Val-Borro et al. 2013; Lim et al. 2014) and morphology (Vasundhara et al. 2007) by means of both on-ground and space observations. This comet was cut from the "science first" approach list due to its retrograde orbit ($i$ = 99.6°), but as it is only slightly retrograde it would have an encounter velocity close to the middle of the range of possible values for CI, and should not be rejected on that basis. This again illustrates that in the real target selection we must consider all available information about a comet.

## 6. CONCLUSIONS AND FUTURE PERSPECTIVES

The exercise performed on the "historical database" and the case study of comet C/2001 Q4 (NEAT) allows us to highlight which elements will be pivotal to investigate and characterise the putative mission target(s), and make decisions on priorities, in the case where there are multiple comets that CI could reach. The requirements of the mission are complex, which is demonstrated by the fact that different approaches to shortlisting result in different preferred targets, but essentially boil down to a preference for the most pristine comet we can reach, expected to be a DNC rather than a returning LPC, and a higher activity body if possible.

The two approaches of the prioritisation exercise do help us identify critical parameters for choosing a target for CI. In practice, real target selection will begin by studying whether or not newly discovered comets are approximately feasible, and those that cannot be reached will not be considered further. The feasibility of individual comets will depend on the final amount of $\Delta v$ available to the mission, which will be known after launch, commissioning, and insertion into the L2 halo orbit (uncertainties on the available launch mass mean that the CI fuel tanks are over-sized for the 600 m/s minimum, with the expectation that any remaining mass margin at launch will be used to top up fuel and maximise $\Delta v$). The exercise revealed a clear preference for more pristine types, so any newly discovered comet with early indication of being a DNC would be a strong contender. While slower encounters are generally preferred, ruling out all retrograde comets removes many otherwise feasible targets, especially those that are not strongly retrograde. It would be better, when considering only feasible targets, to use the actual encounter velocity to highlight particularly favourable (or unfavourable) targets, which will be known as soon as possible intercept trajectories are calculated.

We will have sufficient characterisation of the orbit soon after discovery to select more pristine dynamical classes, in case a rapid decision is needed, and a more reliable DNC/returning assessment after about a year. We are unlikely to have a direct measurement of gas production rates before a decision would need to be made, and even in the case that we have detected CO or $CO_2$, models will be required to extrapolate to total (probably water dominated) gas production at the encounter near 1 au. We will likely have to rely on the (reasonable) assumption that in most cases water production rate correlates with total brightness, and a brighter comet will have a higher gas production rate, and on extrapolation of early brightness evolution to lower heliocentric distance. A better understanding of typical comet behaviours (brightness evolution and any clues about which comets are more likely to have atypical dust-to-gas ratios) will clearly benefit the process by giving us more confidence in these assumptions; this motivates further study of the evolution of distant comets in the years before CI's launch. The prioritisation exercise identified the following measurements and activities that should be made to enable good decisions to be made:

- **Early detection**. The further (in heliocentric distance) the discovery, the more time is available for astrometric, photometric and physical characterisation. LSST will be pivotal for this. Initial orbit fits from discovery observations will be good enough to reveal potentially interesting targets, and discern between prograde and retrograde orbits, for example. Any potential interstellar target will be flagged upon discovery – this will immediately give the start for astrometric and possibly physical observations while the possibility of reaching such a target is considered (this is unlikely for typical warning times and CI's available fuel; see Sánchez & Snodgrass 2025). For more likely Solar System comet targets a ~1-year arc of observations is then needed to evaluate if a LPC coming from the Oort cloud is "new" or "returning".
- **Characterise the nucleus' properties at large distances**. Since the visual magnitude beyond ~10 au is expected to be quite high, the largest telescopes will be needed to try to obtain data on its nucleus, in particular:
    - Nucleus size: Larger comets tend to have higher activity levels (due to their larger surface area). The smallest comets are also more prone to disrupt. Size is prone to degeneracy with active surface area, therefore needs to be deeply investigated

taking into account that reliable information on nucleus size is difficult to obtain solely from ground-based observations when a coma is already present.
    - Nucleus rotation: Comets that are rotating fast are more likely to split due to spin-up beyond the point that self-gravity and centrifugal forces balance (Jewitt 2022).
- **Monitor the coma (dust and gas composition and production), starting as soon as possible**. A regular and continuous monitoring will be needed, to obtain data on early evolution of coma properties, in particular:
    - Evolution of dust production rate.
    - Evolution of gas production rate, where possible.
- **Model the brightness evolution** to infer data on the expected comet's activity at 1 au (close to the encounter). It will be important to find out, e.g., if and at what heliocentric distance jumps/discontinuities might take place. It will be necessary to understand and constrain the activity drivers, which is not so easy. Several authors started to tackle this question, e.g. Gkotsinas et al. (2024) suggest that drivers may be different between JFCs and LPCs, in particular new comets are more likely to be $CO_2$ driven and returning comets are CO driven instead, which seem to agree with A'Hearn et al. (1995), Harrington-Pinto et al. (2022) and Lacerda et al. (2025).

Following the successful first light of the Vera C. Rubin observatory, and with the imminent start of the LSST, we can soon expect to discover comets at large heliocentric distances with some regularity. Whether or not comets are typically active enough for discovery already beyond 10 au, and the true rate of discovery of such comets by LSST, will soon be known. These are perhaps the most fundamental parameters that will drive target selection for CI: if the discovery of comets beyond 10 au remains a rarity in the LSST era then it is perhaps unlikely that the CI mission will have a lot of choices in target selection and will have to simply pick the first possible comet. If LSST finds many distant comets, the exercise described in this paper has given us an insight into the key parameters we need to measure in order to make an optimal choice.

**Acknowledgements**

The authors acknowledge the support of the ESA project team and of the industrial consortium developing the Comet Interceptor mission. We thank the two anonymous referees for comments that helped us improve this manuscript. CS and AD acknowledge funding from the UK Space Agency. EME and CT acknowledge the Italian Space Agency (Contract ASI-INAF n. 2023-14-HH.0). JPS acknowledges support from Centre National d'Études Spatiales. LI was supported by the Italian Space Agency (ASI) within the ASI-INAF agreements I/024/12/0 and 2020-4-HH.0. JDK acknowledges support from BELSPO through ESA PEA 4000139830. AGL was supported by CNES (mission Comet Interceptor). PH was supported by CNES APR. RK acknowledges partial support by project KP-06-D002/3 "CLIC – Cometary Life Cycle," carried out under the PROMYS (Promotion of Young Scientists) component of the Swiss–Bulgarian Research Programme and supported by the Bulgarian National Science Fund. LML acknowledges grants PID2021-126365NB-C21 and Severo Ochoa grant CEX2021-001131-S MICIU/AEI/ 10.13039/501100011033. The views expressed in this article are those of the authors and do not reflect the official policy or position of the U.S. Naval Academy, Department of the Navy, the Department of Defense, or the U.S. Government.

**REFERENCES**

A'Hearn M.F., Millis R.C., Schleicher D.O., Osip D.J., Birch P.V., 1995, *The ensemble properties of comets: Results from narrowband photometry of 85 comets, 1976-1992*, Icarus 118, 2, 223

Bair A.N., Schleicher D.G., Knight M.M., 2018, *Coma Morphology, Numerical Modeling, and Production Rates for Comet C/Lulin (2007 N3)*, AJ 156, 159

Bauer J.M., Grav T., Fernández Y.R. et al., 2017, *Debiasing the NEOWISE Cryogenic Mission Comet Populations*, AJ 154, 2, id. 53

Bergman S., Miyake Y., Kasahara S., Johansson F. L., Henri P., 2023, *Spacecraft Charging Simulations of Probe B1 of Comet Interceptor during the Cometary Flyby*, ApJ 959, 2, id. 138

Betzler A.S., de Sousa O.F., Diepvens A., Bettio T.M., 2020, *BVR photometry of comets 63P/Wild 1 and C/2012 K1 (PANSTARRS)*, ApSS 365, 6, id. 102

Biver N., Bockelée-Morvan D., Crovisier J. et al., 2007, *Submillimetre observations of comets with Odin: 2001 2005*, Planet. Space Sci. 55, 9, p. 1058

Biver N., Bockelée-Morvan, D., Colom P. et al., 2009, *Periodic variation in the water production of comet C/2001 Q4 (NEAT) observed with the Odin satellite*, A&A 501, 1, 359

Biver N., Moreno R., Bockelée-Morvan D. et al., 2016, *Isotopic ratios of H, C, N, O, and S in comets C/2012 F6 (Lemmon) and C/2014 Q2 (Lovejoy)*, A&A 589, A78

Biver N., Bockelée-Morvan D., Handzlik B. et al., 2024, *Chemical composition of comets C/2021 A1 (Leonard) and C/2022 E3 (ZTF) from radio spectroscopy and the abundance of HCOOH and HNCO in comets*, A&A 690, A271

Blum J., Gundlach B., Krause M. et al., 2017, *Evidence for the formation of comet 67P/Churyumov-Gerasimenko through gravitational collapse of a bound clump of pebbles*, MNRAS 469, S755

Bodewits D., Villanueva G.L., Mumma M.J. et al, 2011, *Swift-UVOT Grism Spectroscopy of Comets: A First Application to C/2007 N3 (Lulin)*, AJ 141, 1, id.12

Bortle J.E., 1991, *Post-Perihelion Survival of Comets with Small q*, International Comet Quarterly 13, 89

Bufanda E. Meech K. Kleyna J. et al., 2023, *TNO or Comet? The Search for Activity and Characterization of Distant Object 418993 (2009 MS9)*, The Planetary Science Journal 4, id.2

CI Project Team, 2020, *Comet Interceptor–Mission Requirements* Document, ESA, ESACOMET-SYS-RS-001.

Cochran A.L., Barker E.S., Gray C.L., 2012, *Thirty years of cometary spectroscopy from McDonald Observatory*, Icarus 218, 1, 144

Combi M. R., Mäkinen T. T., Bertaux, J. -L., Quémerais E., Ferron S., 2019, *A survey of water production in 61 comets from SOHO/SWAN observations of hydrogen Lyman-alpha: Twenty-one years 1996-2016*, Icarus 317, 610

Combi M. R., Shou Y., Mäkinen T. et al., 2021, *Water production rates from SOHO/SWAN observations of six comets: 2017-2020*, Icarus 365, id. 114509

Cremonese G., Fulle M., Cambianica P. et al., 2020, *Dust Environment Model of the Interstellar Comet 2I/Borisov*, ApJL 893, L12

Crovisier J., Colom P., Gérard E., Bockelée-Morvan D., Bourgois G., 2002, *Observations at Nançay of the OH 18-cm lines in comets. The data base. Observations made from 1982 to 1999*, A&A 393, p- 1053

De Keyser J., Edberg, N. J. T., Henri P. et al., 2024a, *Optimal choice of closest approach distance for a comet flyby: Application to the Comet Interceptor mission*, Planet. Space Sci. 256, id. 106032

De Keyser J., Edberg, N. J. T., Henri P. et al., 2024b, *In situ plasma and neutral gas observation time windows during a comet flyby: Application to the Comet Interceptor mission*, Planet. Space Sci. 244, id. 105878

de Val-Borro M., Küppers M., Hartogh P. et al., 2013, *A survey of volatile species in Oort cloud comets C/2001 Q4 (NEAT) and C/2002 T7 (LINEAR) at millimeter wavelengths*, A&A 559, A48

Dello Russo N., Kawakita H., Vervack R.J., Weaver H.A., 2016, *Emerging trends and a comet taxonomy based on the volatile chemistry measured in thirty comets with high-resolution infrared spectroscopy between 1997 and 2013*, Icarus 278, p. 301

Edberg, N. J. T., Eriksson A. I., Vigren E. et al., 2024, *Scale size of cometary bow shocks*, A&A 682, A51

ESTEC Concurrent Design Facility (2019), Comet Interceptor CDF Study Report: CDF-201(C), ESA.

Farnham T., 2021, *Comet C/2014 UN271 (Bernardinelli-Bernstein) exhibited activity at 23.8 au*, ATel 14759, 1

Filacchione G., M. Ciarniello M., Fornasier S., Raponi A., 2024, *Comet Nuclei Composition and Evolution*, in "Comets III", Meech K. J., Combi M. R., Bockelée-Morvan D., Raymond S.N., & Zolensky M.E., Eds., Univ. of Arizona Press, pp. 315-360

Fink U., 2009, *A taxonomic survey of comet composition 1985-2004 using CCD spectroscopy*, Icarus 209, 1, 311

Friedel D.N., Remijan A.J., Snyder L.E. et al., 2005, *BIMA Array Detections of HCN in Comets LINEAR (C/2002 T7) and NEAT (C/2001 Q4)*, ApJ 630, 1, 623

Fulle M., Lazzarin M., La Forgia F. et al., 2022, *Comets beyond 4 au: How pristine are Oort nuclei?*, MNRAS 513, 5377

Goetz C., Koenders K., Hansen K.C., et al., 2016, *Structure and evolution of the diamagnetic cavity at comet 67P/Churyumov–Gerasimenko*, MNRAS 462, S459

Gkotsinas A. Nesvorný D. Guilbert-Lepoutre A. Raymond S.N., Kaib N., 2024, *On the Early Thermal Processing of Planetesimals during and after the Giant Planet Instability*, Planet. Space Journal 5, 11, id.243

Groussin O., Lamy P.L., Jorda L., 2010, *The nucleus of comet C/1983 H1 IRAS-Araki-Alcock*, Planet. Space Sci. 58, 6, 904

Guzik, P., Drahus, M., Rusek, K., Waniak, W., Cannizzaro, G., Pastor-Marazuela, I., 2019, *Initial characterization of interstellar comet 2I/Borisov*, Nat. Ast. 4, 53

Harrington-Pinto O., Womack M. Fernandez Y.R., Bauer J., 2022, *A Survey of CO, $CO_2$, and $H_2O$ in Comets and Centaurs*, Planetary Science Journal 3, 11, id. 247

Holt C. E., Knight M. M., Kelley M. S. P. et al., 2024, *Brightness Behavior of Distant Oort Cloud Comets*, Planet. Science Journal 5, 12, id. 273

Holt C.E. & Snodgrass C., 2025, *Empirical Model Improves Future Brightness Predictions for Distant Long-Period Comets*, Planet. Science Journal, submitted

Honkova M., Tichy M., Ticha J. et al., 2013, *Comet C/2013 US10 (Catalina)*, Minor Planet Electronic Circ., No.2013-V31

Hui M.-T., Farnocchia D., Micheli M., 2019, *C/2010 U3 (Boattini): A Bizarre Comet Active at Record Heliocentric Distance*, AJ 157, 162

Inno L., Scuderi M., Bertini I., et al., 2025, *How much earlier would LSST have discovered currently known long-period comets?*, Icarus 429, id. 116443

Ivanova A.V., Korsun P.P., Borisenko S.A., Ivashchenko Y.N., 2013, *Spectral studies of comet C/2001 Q4 (NEAT)*, So. Sys. Res. 47, 2, 71

Jewitt D., 2022, *Destruction of Long-period Comets*, AJ 164, 4, id. 158

Jewitt D., Agarwal J., Hui M.-T. et al., 2019a, *Distant Comet C/2017 K2 and the Cohesion Bottleneck*, AJ 157, 65

Jewitt D., Kim Y., Luu J. Graykowski A., 2019b, *The Discus Comet: C/2014 B1 (Schwartz)*, AJ 157, 103

Jewitt D. & Seligman D.Z. 2023, *The Interstellar Interlopers*, Annu. Rev. Astron. Astrophys. 61, 197

Jones G., Snodgrass C., Tubiana C. et al., 2024, *The Comet Interceptor Mission*, Space Sci. Review 220, 1, id.9

Jorda L., Crovisier J., Green D.W.E., 2008, *The Correlation Between Visual Magnitudes and Water Production Rates*, ACM 2008, held July 14-18, 2008 in Baltimore, Maryland. LPI Contribution No. 1405, paper id. 8046.

Kawakita H. Watanabe J., Furusho R., Fuse T., Boice D.C., 2005, *Nuclear Spin Temperature and Deuterium-to-Hydrogen Ratio of Methane in Comet C/2001 Q4 (NEAT)*, ApJ 623, 1, L49

Kawakita H., Dello Russo N., Furusho R. et al., 2006, *Ortho-to-Para Ratios of Water and Ammonia in Comet C/2001 Q4 (NEAT): Comparison of Nuclear Spin Temperatures of Water, Ammonia, and Methane*, ApJ 643, 2, 1337

Keller H. U., Delamere W. A., Reitsema H. J., Huebner W. F., Schmidt H. U., et al. 1987, *Comet P/Halley's nucleus and its activity*, A&A 187, 1-2, 807

Kidger M., 2023, *Comet Interceptor: A note on target selection in relation to the reference mission scenario*, ESA-COMET-SCI-TN-001

Koenders C., Glassmeier K. -H., Richter I., Motschmann U., Rubin M., 2013, *Revisiting cometary bow shock positions*, Planet. Space Sci. 87, 85

Krolikowska M. & Dybczynski P.A., 2020, The catalogue of cometary orbits and their dynamical evolution, A&A 640, A97

Lacerda P., Guilbert-Lepoutre A., Kokotanekova R. et al., 2025, *Secular brightness curves of 272 comets*, A&A 697, A210

Lamy P.L., Toth I., Fernández Y.R., Weaver H.A., 2004, *The sizes, shapes, albedos, and colors of cometary nuclei*, in "Comets II", M.C. Festou, H.U. Keller, and H.A. Weaver (eds.), University of Arizona Press, Tucson

Langland-Shula L.E. & Smith G.H., 2011, *Comet classification with new methods for gas and dust spectroscopy*, Icarus 213, 1, 280

Lecacheux A., Biver N., Crovisier J. et al., 2003, *Observations of water in comets with Odin*, A&A 402, L55

Levison H.F., 1996, *Comet Taxonomy*, in "Completing the Inventory of the Solar System", Astronomical Society of the Pacific Conference Proceedings, volume 107, T.W. Rettig and J.M. Hahn, Eds., pp. 173-191.

Lim Y.-M., Min K.-W., Feldman P.D., Han W., Edelstein J., 2014, *Far-ultraviolet Observations of Comet C/2001 Q4 (NEAT) with FIMS/SPEAR*, ApJ 781, 2, id. 80

Lippi M., Villanueva G. L., Mumma M. J. et al., 2020, *New Insights into the Chemical Composition of Five Oort Cloud Comets after Re-analysis of Their Infrared Spectra*, Astron. J. 159, 4, id. 157

Lippi M., Villanueva G.L., Mumma M.J., Faggi S., 2021, *Investigation of the Origins of Comets as Revealed through Infrared High-resolution Spectroscopy I. Molecular Abundances*, AJ 162, 2, id.74

Lippi M., Vander Donckt M., Faggi S. et al., 2023, *The volatile composition of C/2021 A1 (Leonard): Comparison between infrared and UV-optical measurements*, A&A 676, A105

Lis D. C., Bockelée-Morvan D. Güsten R. et al., 2019, *Terrestrial deuterium-to-hydrogen ratio in water in hyperactive comets*, A&A 625, L5

Lister T., Kelley M.S.P., Holt C.E. et al., 2022, *The LCO Outbursting Key Project: Overview and Year 1 Status*, Planet. Science Journal 3, 173

Marschall, R., Zakharov, V., Tubiana, C., et al., 2022, *Determining the dust environment of an unknown comet for a spacecraft flyby: The case of ESA's Comet Interceptor mission*, A&A 666, A151

Marsden B., 2002, *COMET C/2001 Q4 (NEAT)*, Minor Planet Electronic Circ. No. 2002-A87

Marsden B., 2004, *COMET C/2001 Q4 (NEAT)*, Minor Planet Electronic Circ. No. 2004-J37.

Mazzotta Epifani E., Dall'Ora M., Di Fabrizio L. et al., 2010, *The activity of comet C/2007 D1 (LINEAR) at 9.7 AU from the Sun*, A&A 513, A33

McGaha J.E., Holvorcem P.R., Schwartz M., Jones G.R., Young J., 2003, *Comet C/2003 T4 (LINEAR)*, IAU Circ., No. 8224, #1

McNaught R.H., Young J., Guido E., Sostero G., 2008, *Comet C/2008 A1 (McNaught)*, IAU Circ., No. 8909, #3

Meech K. J. & Svoren J., 2004, *Using cometary activity to trace the physical and chemical evolution of cometary nuclei*, in "Comets II", M. C. Festou, H. U. Keller, and H. A. Weaver (eds.), University of Arizona Press, Tucson, 745 pp., p.317-335

Meech K.J., Weryk R. Micheli M. et al., 2017, *A brief visit from a red and extremely elongated interstellar asteroid*, Nature 552, 7685, 378

Milam S.N., Remijan A.J., Womack M. et al., 2006, *Formaldehyde in Comets C/1995 O1 (Hale-Bopp), C/2002 T7 (LINEAR), and C/2001 Q4 (NEAT): Investigating the Cometary Origin of $H_2CO$*, ApJ 649, 2, 1169

Opitom C., Jehin E., Manfroid J., et al., 2015a, *TRAPPIST monitoring of comet C/2012 F6 (Lemmon)*, A&A 574, A38

Opitom C., Jehin E., Manfroid J., et al., 2015b, *TRAPPIST photometry and imaging monitoring of comet C/2013 R1 (Lovejoy): Implications for the origin of daughter species,* A&A 584, A121

Opitom C., Guilbert-Lepoutre A., Jehin E. et al., 2016, *Long-term activity and outburst of comet C/2013 A1 (Siding Spring) from narrow-band photometry and long-slit spectroscopy*, A&A 589, A8

Perry R., *Mission Analysis of Intercepting a Dynamically New Comet from a Halo Orbit Launch around the Sun–Earth L2 Libration Point*, MSc in Astronautics and Space Engineering Thesis, Cranfield University, 2019.

Pravdo S.H., Helin E.F., Lawrence K.J. et al, 2001, *Comet C/2001 Q4 (NEAT)*, IAU Circ., No. 7695, #1

Remijan A.J., Friedel D.N., de Pater I. et al., 2006, *A BIMA Array Survey of Molecules in Comets LINEAR (C/2002 T7) and NEAT (C/2001 Q4)*, ApJ 643, 1, 567

Robinson J.E., Malamud U., Opitom C., Perets H., Blum J., 2024, *A link between the size and composition of comets*, MNRAS 531, 859

Rousselot P., Jehin, E., Manfroid J., Hutsemékers D. 2012, *The $^{12}C_2/^{12}C^{13}C$ isotopic ratio in comets C/2001 Q4 (NEAT) and C/2002 T7 (LINEAR)*, A&A 545, A24

Sánchez J.P., Morante, D., Hermosin, P., et al. 2021, *ESA F-Class Comet Interceptor: Trajectory design to intercept a yet-to-be-discovered comet*, Acta Astronautica 188, 265

Sánchez J.P., Snodgrass C., Küppers M., Mazzotta Epifani E., et al., 2024, *Comet Interceptor: An ESA Mission to a Yet Unidentified Target*, 75th International Astronautical Congress (IAC), Milan, Italy, 14-18 October 2024

Sánchez J.P., & Snodgrass C., 2025, *Analysis of Trajectories to 3I/ATLAS with a Comet Interceptor-like Spacecraft*, Research Notes of the AAS 9, id.207

Schleicher D. G., Woodney L. M., Birch P. V., 2002, *Photometry and Imaging of Comet C/2000 WM1 (Linear)*, Earth, Moon & Planets 90, 1, 401

Schleicher D. G. & Osip D. J., 2002, *Long- and Short-Term Photometric Behavior of Comet Hyakutake (1996 B2)*, Icarus 159, 1, 210

Sekanina Z., 2019, preprint, https://arxiv.org/abs/1903.06300

Seligman, D.Z., Micheli, M., Farnocchia, D., et al., 2025, *Discovery and Preliminary Characterization of a Third Interstellar Object: 3I/ATLAS*, ApJL 989, L36

Shinnaka Y., Kawakita H., Kobayashi H., Boice D.C., Martinez S.E., 2012, *Ortho-to-para Abundance Ratio of Water Ion in Comet C/2001 Q4 (NEAT): Implication for Ortho-to-para Abundance Ratio of Water*, ApJ 749, 2, id. 101

Snodgrass C., A'Hearn M. F., Aceituno F. et al., 2017, *The 67P/Churyumov-Gerasimenko observation campaign in support of the Rosetta mission*, Phil. Trans. R. Soc. A 375, 20160249

Snodgrass, C., Feaga, L., Jones, G. H., Küppers, M., Tubiana, C., & Dotson, R. (2024). *Past and Future Comet Missions*. in "Comets III", Meech K. J., Combi M. R., Bockelée-Morvan D., Raymond S.N., & Zolensky M.E., Eds., Univ. of Arizona Press, pp. 155–192

Snodgrass C., Holt C.E., Kelley M.S.P. et al., 2025, *First JWST spectrum of distant activity in long-period comet C/2024 E1 (Wierzchos)*, MNRAS 541, 1, L8

Tozzi G.P., H. Böhnhardt H., Lo Curto G., 2003, *Imaging and spectroscopy of comet C/2001 Q4 (NEAT) at 8.6 AU from the Sun*, A&A 398, L41

Vasundhara R., Chakraborty P., Muneer S., Masi G., Rondi S., 2007, *Investigations of the Morphology of Dust Shells of Comet C/2001 Q4 (NEAT)*, AJ 133, 2, 612

Vigren E., Eriksson A. I., Edberg N. J. T., Snodgrass C., 2023, *A potential aid in the target selection for the comet interceptor mission*, Planet. Space Sci. 237, id. 105765

Whipple F.L., 1992, *A volatility index for comets*, Icarus 98, 1, 108

Wooden D.H., Woodward C.E., Harker D.E., 2004, *Discovery of Crystalline Silicates in Comet C/2001 Q4 (NEAT)*, ApJ 612, 1, L77